 \documentstyle[prl,aps,epsfig,twocolumn]{revtex}

\newlength{\textwidthm}
\begin{document}
\title{Dark-State Polaritons in Electromagnetically Induced Transparency}
\author{M.~Fleischhauer$^1$ and M.~D.~Lukin$^2$}
\address{$^1$ Sektion Physik, Ludwig-Maximilians-Universit\"at M\"unchen,
Theresienstr. 37, D-80333 M\"unchen, Germany}
\address{$^2$ ITAMP, Harvard-Smithsonian Center for Astrophysics,
                Cambridge, MA~~02138 }
\date{\today}
\maketitle
\begin{abstract}
We identify form-stable coupled excitations of light and matter 
(``dark-state polaritons'')
associated with the propagation of quantum fields in 
Electromagnetically Induced Transparency. The properties of the
dark-state polaritons such as the group velocity
are determined by the mixing angle between light and matter components 
and can be controlled by an external coherent field as the pulse propagates.  
In particular, light pulses can be decelerated and ``trapped'' in which case 
their shape and quantum state are mapped onto metastable 
collective states of matter.  Possible applications of this reversible  
coherent-control technique are discussed. 
\end{abstract}


\nobreak
Dark resonances and electromagnetically induced transparency 
(EIT) \cite{EIT,dark} can be used to make a resonant, optically opaque medium 
transparent by 
means of quantum interference.
Associated with the induced transparency is a dramatic modification of the 
refractive properties of the media. These can result, for instance, 
in very slow  group velocities \cite{Hau99}. 
In the present contribution we study 
the propagation of quantum fields in EIT media. 
We demonstrate the existence of 
formstable
quantum excitations associated with such propagation, which we 
term  ``dark-state polaritons''. 
The polaritons are coherent superpositions of photonic and Raman-like matter 
branches.
We show that their group
velocity is directly related to
the ratio of the two contributions. This ratio can be externally controlled by 
adiabatically changing a coherent control field as the pulse propagates. 
In particular, dark-state polaritons  
can be stopped and re-accelerated
in such a way that their shape and quantum state are preserved. In this 
process the quantum state of light is ideally transfered to 
collective atomic excitations and vise versa.

The possibility to coherently control the propagation of quantum light pulses
via dark-state polaritons opens up 
interesting applications involving the generation of non-classical states of
atomic ensembles (in squeezed or 
entangled states), reversible quantum memories for light waves 
\cite{Fleischhauer00,Lukin00,Polzik_pre}, and 
high resolution spectroscopy \cite{wine}. Furthermore, 
the combination of the
present technique with studies on few-photon nonlinear optics 
\cite{Hau99b,Imamoglu,LukinNLO,zibrov99,NLO_rev}
can be  used, in principle, for processing of 
quantum information stored in collective excitations of matter. 
Finally, the present technique 
may provide an interesting tool to study
quantum scattering phenomena in systems  involving 
coherent cold collisions. In this regard the present work opens 
a link between nonlinear optics for light waves and nonlinear 
atom optics. E.g.~an interaction (or entanglement) between light waves 
can be induced by a collisional 
interaction of atoms (e.g. s-wave scattering); 
alternatively an interaction between
atoms can be induced via optical nonlinearities.

We consider a medium consisting of $\Lambda$-type 3-level atoms with 
two meta-stable lower states as shown in Fig.~1. A quantum field
described by the slowly-varying dimensionless operator 
\begin{eqnarray}
\hat E(z,t)=
\sum_k \, a_k(t) \, {\rm e}^{{ i} kz}
\,{\rm e}^{-{ i}\frac{\nu}{c}(z-ct)} 
\end{eqnarray}
couples resonantly
the transition between the ground state $|b\rangle$ and
the excited state $|a\rangle$. $\nu=\omega_{ab}$ is the carrier frequency 
of the optical field.
The upper level $|a\rangle$ is furthermore coupled to the stable state 
$|c\rangle$ via a coherent control field
with the slowly-varying, real Rabi-frequency $\Omega(t)$.
For the purposes of the present discussion 
the external field can be treated classically.  
We assume that initially (i.e before the quantum pulse arrives) all atoms are 
in their ground states $|b_j\rangle$. 
To describe the quantum properties of the medium, we use collective,
slowly varying 
atomic operators, appropriately averaged over small but macroscopic volumes
containing $N_z\gg 1$ particles at position $z$, 
\begin{equation}
\hat\sigma_{\alpha\beta}(z,t) = {1 \over N_z} 
\sum_{j=1}^{N_z} |\alpha_j\rangle\langle\beta_j|\, 
{\rm e}^{-{ i}\omega_{\alpha\beta} t}.  
\end{equation}
The interaction between light and atoms is governed by the Hamiltonian
\begin{equation}
{\hat V} = -N
\int\!\! \frac{{\rm d} z}{L} 
\left( \hbar g \sum_k a_k {\rm e}^{{ i} k z} \hat\sigma_{ab}(z) + 
\hbar\Omega \hat\sigma_{ac}(z) \right)
+ h.c.
\label{ham}
\end{equation}
Here $g=\wp\sqrt{\frac{\nu}{2\hbar\epsilon_0 V}}$ is the atom-field 
coupling constant with $\wp$ being the dipole moment of the 
$a-b$ transition and $V$ the quantization volume. $
N$ is the number of atoms in this volume and $L$ its length in
$z$ direction.

The evolution of the Heisenberg operator corresponding to the 
optical field can be described in slowly varying 
amplitude approximation by the propagation equation

\begin{eqnarray}
\left(\frac{\partial}{\partial t}+c\frac{\partial}{\partial z}\right)
\hat E(z,t)= { i} g N\, \hat\sigma_{ba}(z,t).\label{field}
\end{eqnarray}
The atomic evolution is governed by a set of 
Heisenberg-Langevin equations
\begin{eqnarray}
\frac{\partial}{\partial t} \hat \sigma_{\mu\nu} = - \gamma_{\mu\nu} 
\sigma_{\mu\nu} + 
{i \over \hbar} 
[{\hat V},{\hat \sigma}_{\mu\nu}] + F_{\mu\nu},
\label{s_cb}
\end{eqnarray}
where $\gamma_{\mu\nu}$ are the transversal decay rates and $\hat F_{\mu\nu}$
are $\delta$-correlated Langevin noise operators. 

We now assume that the Rabi-frequency of the quantum field
is initially much smaller than $\Omega$ and that the
number of photons in the input pulse is much less than the
number of atoms. We will show that the Rabi-frequency of the quantum
field will then be much smaller than $\Omega$ at all times.
In such a case the atomic 
equations can be treated perturbatively in $\hat E$. 
In zeroth order only $\hat\sigma_{bb}={\bf 1}$ is different from zero
and in first order 
one finds 
\begin{eqnarray}
\hat\sigma_{ba}&=&-\frac{ i}{\Omega(t)} \frac{\partial}{\partial t}
\hat \sigma_{bc},\\
\hat\sigma_{bc}&=&
-\frac{g \hat E}{\Omega} -\frac{ i}{\Omega}\left[
\left(\frac{\partial}{\partial t}+\gamma_{ba}\right)\left(-\frac{i}{\Omega}
\frac{\partial}{\partial t}\hat\sigma_{bc}\right)+\hat F_{ba}\right].
\label{s_cb_2}
\end{eqnarray}
In the above equations we disregarded a (small) decay of the Raman 
coherence ($\gamma_{bc}$). 

The propagation equations simplify considerably if we assume
a sufficiently slow change of $\Omega$, i.e. adiabatic conditions
\cite{Hau99b,LukinNLO}.
Introducing a normalized time $\tilde t=t/T$
where $T$ is a characteristic time scale 
and expanding the r.h.s. of (\ref{s_cb_2}) in powers of $1/T$
we find in lowest non-vanishing order  
\begin{equation}
\hat\sigma_{bc}(z,t)
=-g\frac{\hat E(z,t)}{\Omega(t)}.\label{s_cb_ad}
\end{equation}
Note that 
$\langle \hat F_x(t) \hat F_y(t')\rangle \sim \delta(t-t')
= \delta(\tilde t-\tilde t')/T$.
Thus in the perturbative and adiabatic limit
the propagation of the quantum light pulse is governed by the equation
\begin{eqnarray}
\left(\frac{\partial}{\partial t}+c\frac{\partial}{\partial z}\right)
\hat E(z,t)= -\frac{g^2 N}{
\Omega(t)}\, \frac{\partial}{\partial t}\frac{ \hat E(z,t)}{\Omega(t)}.
\label{field_ad}
\end{eqnarray}
If $\Omega$ is constant, the term on the r.h.s. 
simply leads to a modification
of the group velocity of the quantum field according to 
$v_g=c/(1+{\frac{g^2 N}{\Omega^2}})$.
In the general case the field equation of motion will acquire
an additional term  proportional to $(\dot\Omega/\Omega)\, \hat E$ 
which describes
reversible changes in quantum amplitudes due to stimulated Raman scattering. 

One can obtain 
a very simple solution of eq.(\ref{field_ad})
by introducing a new quantum field
$\hat \Psi(z,t)$ via the canonical transformation
\begin{eqnarray}
&&\hat\Psi(z,t)=\cos\theta(t)\, \hat E(z,t) - \sin\theta(t)\, \sqrt{N}\,
\hat \sigma_{bc}(z,t), \\
&&\cos\theta(t) = {\Omega(t) \over \sqrt{\Omega^2(t) + g^2 N}}, \; \; 
\sin\theta(t) = {g \sqrt{N} \over \sqrt{\Omega^2(t) + g^2 N}}.
\nonumber
\end{eqnarray}
$\hat\Psi$ obeys the following equation of motion
\begin{eqnarray}
\left[\frac{\partial}{\partial t}+c\cos^2\theta(t)
\frac{\partial}{\partial z}\right]\hat\Psi(z,t)=0,
\end{eqnarray}
which describes a shape-preserving propagation with velocity
$v=v_g(t)=c\cos^2\theta(t)$: 
\begin{equation}
\hat\Psi(z,t)=\hat \Psi\left[z- c\int^t_0\!\!\!{\rm d}\tau
\cos^2\theta(\tau),t=0\right].
\label{sol}
\end{equation}

Several interesting properties of the new field should be noted.
First of all, by introducing a plain-wave decomposition 
$\hat\Psi(z,t)=\sum_k \hat\Psi_k(t)\, {\rm e}^{ikz}$ one finds that
the mode operators $\hat\Psi_k$ and $\hat\Psi_k^\dagger$ obey the commutation
relations
\begin{eqnarray}
[\hat\Psi_k, \hat\Psi_{k'}^+] = \delta_{k,k'}\, \Bigl[\cos^2\theta + 
\sin^2\theta\frac{1}{N}\sum_j
({\hat \sigma}_{bb}^j-{\hat \sigma}_{cc}^j)\Bigr].
\end{eqnarray}
In the linear limit considered here, where the number density
of photons is much smaller than the density of atoms,
${\hat \sigma}_{bb}^j \approx 1, {\hat \sigma}_{cc}^j \approx 0$.
Thus the new field possesses bosonic commutation relations and we can
associate with it bosonic quasi-particles (polaritons). 
Furthermore one  immediately verifies that all number states created by
$\hat\Psi_k^\dagger$ are dark-states \cite{dark,Lukin00}:
\begin{eqnarray}
|D_n^k\rangle = \frac{1}{\sqrt{n!}}
\Bigl(\hat\Psi_k^\dagger\Bigr)^n |0\rangle |b_1 ...b_N\rangle,
\label{dark} 
\end{eqnarray}
where $|0\rangle$ denotes the field vacuum.
In particular, the states $|D_n^k\rangle$ do not contain the excited
atomic state and are thus immune to spontaneous emission. Furthermore, 
they are eigenstates of the interaction Hamiltonian with eigenvalue zero, 
${\hat V}\, |D_n^k\rangle  = 0$.
For these reasons we call the quasi-particles ``dark-state polaritons''.

To summarize, we have found a shape-preserving, 
polariton-like superposition $\hat\Psi$ of an electromagnetic field
and collective Raman coherences. 
This excitation is not of soliton type since no special pulse-shape 
or pulse area is required. It is related to the classical 
adiabaton solutions
of pulse-pair propagation in $\Lambda$-type media
\cite{Grobe95,Cerboneschi95,Fleischhauer96} in the limit of one strong and one
weak field. We emphasize however that the field can here be in any
quantum state. In particular it does not need to have a coherent
component with a well defined phase.

One of the most interesting aspects of dark-state polaritons is
the possibility to  coherently control their properties 
by changing $\Omega(t)$. For example, 
by adiabatically rotating $\theta(t)$ from $0$ to $\pi/2$
one can decelerate and stop an input light pulse. 
It is remarkable that in this process pulse shape and quantum 
state of the initial light pulse
are mapped onto collective, metastable states of matter
in which they are stored. Likewise the dark-state polariton can be
re-accelerated to the vacuum speed of light; in this process the stored 
quantum states is transferred back to the field. 
This is illustrated in Fig.~2, where we have shown 
the coherent amplitude of a dark-state polariton which results from an initial
light pulse as well as the corresponding  field and matter
components. 
One recognizes that the pulse shape is preserved and that the 
stopping corresponds to a transfer from field 
to atomic Raman excitations.
Explicitly, the mapping of the quantum states corresponds to the following 
unitary  transformation:
\begin{eqnarray} 
&&\left(\sum_{k,l,m...}\xi_{k,l,m...} a^\dagger_k a^\dagger_l a^\dagger_m...|0\rangle \right)
|b_1...b_N\rangle
\leftrightarrow  \label{state}  \\
&& \left(
\sum_{k,l,m...}\xi_{k,l,m...} \sqrt{N} \sigma_{cb}^k
\sqrt{N} \sigma_{cb}^l \sqrt{N} \sigma_{cb}^m ...|b_1...b_N\rangle \right)
|0\rangle, \nonumber
\end{eqnarray} 
as can be verified using expression (\ref{dark}) for the 
polariton state vectors.

The coherent transfer of quantum states between light and 
matter opens interesting
prospectives for the generation of non-classical atomic ensembles in squeezed
and entangled states, high-precision
spectroscopy with resolution beyond the standard quantum 
limit \cite{wine} as well
as reversible quantum memories. Furthermore, by trapping correlated photons in 
separate media entangled states of separated atomic ensembles 
can be created. With respect to these applications
the present paper is complementary to our earlier studies in which we 
showed that quantum states of light can be mapped onto Dicke-like 
collective states of an EIT medium in an optical resonator
\cite{Fleischhauer00,Lukin00}. The quantum
states of matter generated in the case of the present paper
are more complicated; 
however trapping the light in a traveling-wave geometry does not require 
special shaping of the classical driving pulses (quantum impedance matching), 
which is necessary in a cavity configuration. 

We also note related studies on quantum memories for light 
involving mapping the quantum state of the field onto atoms by 
dissipative absorption   
 \cite{Polzik_pre,Polzik1}. In contrast to these approaches 
the adiabatic passage technique
\cite{STIRAP} used here allows for  a complete and reversible excitation
transfer of arbitrary quantum wavepackets.

Finally, our approach is also 
different from the mechanism suggested recently in \cite{olga}, in which 
``freezing'' of the light pulse in a laboratory frame was proposed
using moving atoms. 

The above analysis  involves a perturbation expansion, an adiabatic
approximation and disregards the decay of Raman coherence.  In what  
follows the validity of these approximations is discussed. 
First of all, we note that making use of (\ref{s_cb_ad}) one finds:
$g^2 {\hat E}^+{\hat E}/ |\Omega|^2 = {\hat \sigma}_{cb} {\hat \sigma}_{bc}$.
I.e., the ratio of the average intensities of quantum and 
control field
is proportional to that of the matter field 
$\langle\hat\sigma_{cc}\rangle$. 
If the initial number of photons in the quantum field is much less
than the number
of atoms,  $\langle\hat\sigma_{cc}\rangle$ is always much smaller than 
unity. Therefore the mean  intensity of the quantum field 
remains
 small compared to that of the control field even when the latter 
is turned to zero. 

In order to check the validity of the adiabatic
approximation we consider the first 
correction to 
$\hat\sigma_{bc}$:
\begin{eqnarray}
\hat\sigma_{bc}\approx 
-\frac{g \hat E}{\Omega} +\frac{ 1}{\Omega}
\left(\frac{\partial}{\partial t}+\gamma_{ba}\right)
\frac{1}{\Omega}\frac{\partial}{\partial t}\frac{g\hat E}{\Omega} + \cdots
\label{correct}
\end{eqnarray}
The non-adiabatic correction in (\ref{correct}) 
leads to a spectral narrowing (pulse spreading) of the quantum 
field due to the finite bandwidth of the transparency window \cite{LukinNLO},
which results in a ``pulse''-matching of quantum and classical control
field \cite{Harris9394,Fleischhauer96}.
Using the adiabatic solution (\ref{sol}), one can 
verify that these corrections are small for propagation 
distances:
\begin{eqnarray}
z \ll z_{max} = {g^2 N \over \gamma_{ab}} \times {L_p^2 \over c},  
\end{eqnarray}
where $L_p$ is the length of the input pulse. Hence, in order to 
trap a pulse with negligible losses, it is required that 
\begin{eqnarray}
{g^2 N L_p \over c\gamma_{ab}} \gg 1. 
\label{adi}
\end{eqnarray}
This condition contains the number of atoms which 
is a signature of collective interactions. It should 
be contrasted to the
strong-coupling condition corresponding to a quantum state transfer in cavity 
QED \cite{QED}. 
We note, in particular,  that in the optically dense medium the adiabatic 
condition (\ref{adi}) is  much easier to implement. 

The effect of the Raman coherence decay can be easily estimated using  
the explicit expression for the generated matter states (\ref{state}). 
It is clear that the collective states containing $n_e$ atomic
excitations will 
dephase at a rate $\gamma_{bc} \, n_e$. Hence, the time of the 
storage should be limited to $ t_s \ll (\gamma_{bc} \, n_e)^{-1}$ to 
avoid  decoherence \cite{Lukin00}.

In the discussion above we have considered the case where the
control field only depends on time. This is valid, for instance, when 
the control field propagates in a direction 
perpendicular to that of the quantum field. In experiments involving hot 
atomic vapors co-propagation is required, however, in order to cancel Doppler 
broadening of the  two-photon transition. In this case 
propagation
effects of the control field need to be considered.
If the quantum field is weak, the control field propagates 
as in free space and thus $\Omega(z,t)=\Omega(t-z/c)$. In this 
case one finds:
\begin{eqnarray}
\left(\frac{\partial}{\partial t}+c
\cos^2\theta(z,t)\frac{\partial}{\partial z}\right)\, 
\frac{\hat E(z,t)}{\Omega(z,t)}=0.
\end{eqnarray}
Since the group velocity is now also
$z$-dependent, trapping of the pulse does not preserve the shape 
exactly. Nevertheless 
it is evident that trapping and a reversible transfer of the quantum  
state from light to atoms are still possible. In experiments, however, 
a more practical approach can be taken in which 
a light pulse enters the medium already with $v_{g}^0 \ll c$.
In such a case retardation of the control field can be ignored and one 
has $\Omega(t-z/c)\approx \Omega(t)$. 
Since the index of refraction is close to unity there will
be no reflection losses at the entrance plane. 
However the polariton pulse becomes spatially compressed
according to $L_p/L_p^0=v_{g}^0/c$, and its amplitude grows 
according to the boundary condition ${\hat \Psi}(0,t) = \sqrt{c/v_g^0}
{\hat E}(0,t)$. In this way, the total energy of the polariton field 
inside the medium is equval to the energy of the light field outside. 
After entering the medium the polaritons can be manipulated 
as discussed above.

In conclusion we have shown that it is possible to control 
the propagation 
of quantum pulses in optically thick $\Lambda$-type media. This coherent 
control mechanism is based on  dark-state polaritons associated with
 EIT. In particular, 
a quantum light pulse can be ``trapped'', in which case its shape
and quantum state are preserved in stationary atomic excitations.
The matter-like polariton can then be re-accelerated
and converted back into a photon pulse. These properties of dark-state 
polaritons can be used for squeezing and entanglement transfer from light to
atoms. Furthermore, we
anticipate interesting applications involving  nonlinear interactions
between such polaritons. 

We thank M.O. Scully  for many stimulating 
discussions. This work was supported by the National Science Foundation.


\def\etal{\textit{et al.}}


\begin{figure}[ht]
\centerline{\epsfig{file=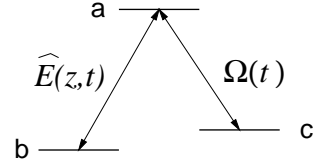,width=4.0cm}}
\vspace*{2ex}
\caption{3-level $\Lambda$-type medium resonantly coupled to a classical
field with Rabi-frequency $\Omega(t)$ and quantum field $\hat E(z,t)$.}
\end{figure}


\begin{figure}[ht]
\centerline{\epsfig{file=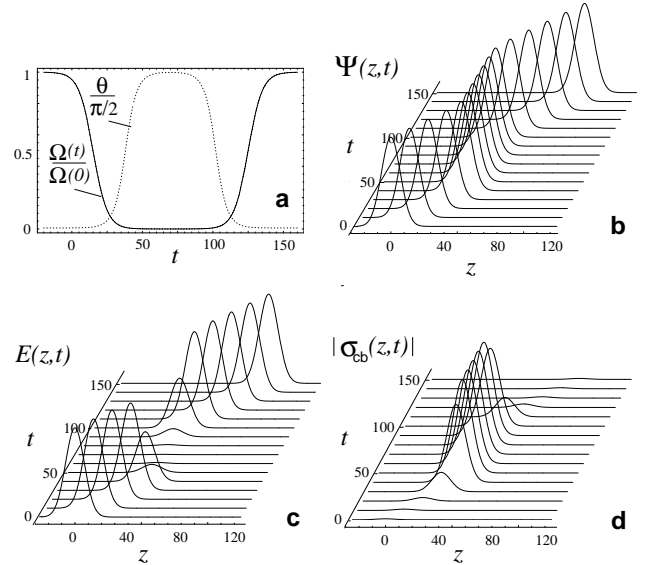,width=9.0cm}}
 \vspace*{2ex}
 \caption{Propagation of a dark-state polariton 
with envelope $\exp\{-(z/10)^2\}$. 
The mixing angle is rotated from $0$ to $\pi/2$ and back
according to $\cot\theta(t)  =$ $ 100(1-0.5 \tanh[0.1(t-15)]$ $
 + 0.5\tanh[0.1(t-125)])$ as shown in (a). 
 The  
coherent amplitude of the polariton $\Psi=\langle\hat\Psi\rangle$ 
is plotted in (b) and the electric field $E=\langle \hat E\rangle
$ and matter components $|\sigma_{cb}|=|\langle\hat \sigma_{cb}\rangle|$
in (c) and (d) respectively. 
 Axes are in arbitrary units with $c=1$. }
\end{figure}

\end{document}